# Population Activity Recovery: Milestones Unfolding, Temporal Interdependencies, and Relationship with Physical and Social Vulnerability

**Flavia Ioana Patrascu[1*], Ali Mostafavi[2]**

[1]Ph.D. Candidate, Urban Resilience.AI Lab, Zachry Department of Civil and Environmental Engineering, Texas A&M University, College Station, TX 77843, USA; e-mail: patrascu.flavia.ioana@gmail.com [*]corresponding author

[2]Full Professor, Urban Resilience.AI Lab Zachry Department of Civil and Environmental Engineering, Texas A&M University, College Station, TX 77843, USA; e-mail: amostafavi@civil.tamu.edu

## Abstract

Understanding sequential community recovery milestones is crucial for proactive recovery planning and monitoring. This study investigates these milestones related to population activities to examine their temporal interdependencies and evaluate the relationship between recovery milestones and physical (residential property damage) and social vulnerability (household income). This study leverages post-2017 Hurricane Harvey mobility data from Harris County to specify and analyze temporal recovery milestones and their interdependencies. The analysis examined four key milestones: return to evacuated areas, recovery of essential and nonessential services, and the rate of home-switch (moving out of residences). Robust linear regression validates interdependencies between across milestone lags and sequences: achieving earlier milestones accelerates subsequent recovery milestones. The study thus identifies six primary recovery milestone sequences. We found that social vulnerability accounted through the median household income level, rather than physical vulnerability to flooding accounted through the property damage extent, correlates with recovery delays between milestones. We studied variations in recovery sequences across lower and upper quantiles of property damage extent and median household income: lower property damage extent and lower household income show greater representation in the "slowest to recover" sequence, while households with greater damage and higher income are predominant in the group with the "fastest recovery sequences". Milestone sequence variability aligns closely with social vulnerability, independent of physical vulnerability. Understanding the variation in recovery sequences, milestone interdependencies, and social vulnerability disparities provides crucial evidence for targeted interventions. This empowers emergency managers to effectively monitor and manage recovery efforts, enabling timely interventions.

# 1 Introduction

Natural weather-related disasters are becoming more frequent and severe, exacerbated by an increasing high-density urbanization and climate change [1]. As cities expand and become more densely populated, the complexity and scale of managing disaster recovery efforts also increases. Extreme weather-related events, such as storms and hurricanes, create significant disruptions to both the natural and built environments [2], especially in densely populated coastal areas [3, 4]. These occurrences require thorough preparation and targeted strategies to mitigate their effects, which can severely disrupt urban social, economic, and physical infrastructure [5, 6]. Addressing post-disaster community recovery involves the interconnected elements of the built environment, infrastructure, and population. In the context of community recovery, rapid assessment allows for timely intervention and supports the allocation of resources where they are most needed to facilitate a return to the pre-disaster state [7, 8].

Understanding and effectively managing the four major phases of disasters—mitigation, preparedness, response, and recovery—has been a critical area of focus especially in the context of earthquakes and weather-related events since the inception of this type of research in the 1980s [9]. Over the following decade, significant efforts have been dedicated to identifying these phases to better cope with the devastating effects of disasters. Some studies have primarily concentrated on the substantial costs linked to physical and human losses, which have escalated due to the exponential growth in population and infrastructure since the early 20th Century [10, 11, 12]. Initially, scholars defined community recovery in terms of economic, housing, quality of life, and emotional recovery [11] and later on to include the housing, business, public services, facilities and people [9].

The standard community disaster recovery literature [13, 14, 15, 16, 17, 18], divides the recovery process into stages of short-term, intermediate, and long-term recovery with qualitative descriptions of recovery activities at each stage. This rather subjective approach to community recovery characterization in the extant literature [19, 20, 21] is particularly problematic in three aspects: (1) lack of specification of critical recovery milestones, as well as quantitative data-driven measures corresponding to milestones to enable proactive monitoring of recovery in different areas to inform recovery implementation and resource allocation; (2) limited understanding of the underlying vulnerability characteristics, whether physical - such as buildings and infrastructure systems - or social - such as population-related characteristics - that influence the spatial and temporal patterns of recovery; and (3) inadequate focus on temporal interdependence among milestones and limited examination of disparities in recovery trajectories among different sub-populations. Departing from the standard approach for designation of the recovery stage, this study determines critical milestones based on the characteristics of human mobility networks related to population activities. Critical recovery milestones are defined as times at which life activities of populations return to their pre-disaster level (or exceed pre-disaster levels through building back better). Accordingly, the recovery trajectory captures the sequence and duration of different critical recovery milestones. The specification and characterization of these critical community recovery times are essential to enable the examination of the trajectory of recovery for sub-populations and the extent to which different recovery milestones are temporally interdependent. These important data-driven insights enable the understanding of community recovery at scales and a level of complexity that has not been investigated before.

The research approach for addressing the study objectives is particularly designed to fill important gaps in the existing literature. First, although a number of recent studies [22, 23, 24, 25] have examined mobility networks and their properties during disasters, the majority of these studies focus on evacuation patterns [26, 27, 28], disruption in mobility [29, 30, 31, 32], and theoretical network resilience properties [31, 33, 34, 35, 36], and also focus mainly on origin-destination networks [37, 38]. Despite the recognition that socio-spatial networks related to population movements and activities play important roles in community resilience and recovery in disasters, little of the existing work has attempted to specify features of community recovery from the topological structure and dynamical properties of these socio-spatial networks. Also, only a limited number of studies have examined population-facility networks (patterns of visitations to points of interest) and their characterization of community resilience and recovery based on the structure and properties of these significant spatial networks. This limitation is particularly problematic, as it has led to the absence of data-driven methods and measures to quantify and proactively monitor the trajectory of recovery milestones at the proper scale needed to inform recovery implementation and resource allocation.

Second, while the existing literature has examined various factors influencing community disaster recovery [9, 39, 40, 41], the current approach to understanding disaster recovery considers recovery as a single milestone rather than a sequence of multiple milestones. Despite the well-documented evidence regarding the presence of recovery disparity, there is limited knowledge regarding the difference in the recovery trajectory (i.e., sequence of recovery milestones) and the effects of community vulnerability (both physical and social vulnerability) on recovery trajectories and disparity. Hence, in this study, we examine the correlation between physical and social vulnerability characteristics and the recovery trajectory of spatial blocks.

Accordingly, the study aims to address the following research questions: (1) To what extent is the temporal unfolding of recovery milestones consistent across different spatial blocks? What are the similarities and differences in the patterns of milestones unfolding? (2) To what extent can previously achieved milestones influence subsequent ones? Is there variability in the time gaps between these milestones? and (3): To what extent do physical and social vulnerability explain variations in recovery milestone sequences across spatial blocks? To what degree do vulnerability characteristics explain disparities in the lags among the milestones across different areas? To answer the research questions, we utilized location intelligence data in the context of the 2017 Hurricane Harvey in Harris County, Texas, to specify population activity recovery milestones as a component of the overall community recovery. We particularly examine the evacuation return, essential activities, nonessential activities, and move-out return, as four critical milestones to gauge the population activity recovery timeline across spatial blocks. Accordingly, we examine variations in the temporal sequence of these milestones, lags among the milestones across different sequences, and disparities in the lags in association with physical and social vulnerability characteristics. Identifying specific milestone sequences can reduce the complexity of uncertainty and explain the previously perceived non-linear nature of community recovery. Studying these sequences and their temporal dependencies can significantly improve the planning and preparedness stages of disaster management. By understanding the typical sequence of recovery milestones, policymakers and emergency planners can develop more effective recovery plans, allocate resources more efficiently, and prioritize critical activities during recovery phases. Moreover, understanding temporal patterns and dependencies improves the predictability of the recovery process, helping to establish accurate expectations for recovery efforts.

To address the research questions of this study, the paper is structured as follows: Section 2 covers the data, Section 3 details the methods, Section 4 analyzes the results, and Section 5 provides concluding remarks.

## 2   Data

### 2.1   Study context

The study region is in Harris County, Texas, during the devastating Hurricane Harvey in 2017, which resulted in unprecedented flooding throughout the region [42, 43]. The third most populous county in the United States, Harris County encompasses Houston, the nation's fourth-largest city. The area's susceptibility to flooding is aggravated by several factors: its flat coastal location with unrestricted development in flood plains, its proximity to the Gulf of Mexico, which brings warm moist air masses, a rapidly expanding population, and ongoing urban development. Hurricane Harvey's impact on the area, coupled with the extensive availability of mobility data for Harris County before and after the landfall, makes it an ideal setting to explore the research questions posed in this study.

### 2.2   Data processing

#### 2.2.1   Recovery milestones

The recovery milestones measurement metrics were calculated using a three-step methodology motivated by the work of Jiang et al. [25]. Similar methodologies were employed in other recent studies to account for population activity fluctuations in disaster settings, including visits to points of interest (POIs) [44, 45, 46],  movements from census block groups (CBG) to POIs [32, 47, 48, 49], and visits between CBGs [28, 48, 50, 51].

Spectus, a commercial company, provides privacy-enhanced anonymous mobility data. It currently collects data from about 15 million users in the United States, ensuring that datasets are representative of population mobility. From each user, Spectus gathers an average of around 100 data points daily, sourced from third-party collaborator apps whose users agree to share their location information. Location intelligence data, which include precise GPS coordinates of trip destinations, comply with the General Data Protection Regulation and the California Consumer Privacy Act frameworks. To protect the privacy and confidentiality of third-party users, Spectus de-identifies and aggregates home locations of users at the CBG level.

Moreover, we utilized the North American Industry Classification System (NAICS) codes to identify essential POIs based on the SafeGraph dataset. SafeGraph's database includes spatial coordinates and identifying information, such as business type, operating hours, and NAICS codes, for over 30 million POIs.

The Microsoft's Building Footprint Data was used to pinpoint the POI locations. This step was necessary to align these locations with the SafeGraph dataset, since the latter did not include this information for the year 2017, which is the focus period of this research.

### 2.2.1.1   CBG-to-POI movements

The daily visits from each CBG to POI were analyzed using Spectus data. This analysis captured the user or household population activity in terms of moving from a home CBG to a POI (e.g., businesses and critical facilities). The fluctuations in visits to POIs were used to account for the recovery milestones described below:

*Essential recovery milestone* was determined using POIs deemed as a requirement for life to continue, such as drug stores, healthcare facilities, grocery stores, and utilities (electric, gas, water, and sanitation facilities)—services that, if disrupted, would cause hardship for the population. For instance, during extreme weather events, individuals should still have access to food, hospitals, and medicine. Furthermore, during service disruptions, such as electricity or water supply outages, people generally prefer to speak with a customer service representative over the phone [52, 53]. If this option is unavailable, they often seek in-person communication. Consequently, movement to such facilities is considered essential [5].

*Non-essential recovery milestone* was calculated using POIs associated with activities such as self-care facilities, retail, recreation, or restaurants, which were considered nonessential for the population survival, but which are a good proxy for assessing population lifestyles and return to normal lifestyle from a disaster situation [54, 55].

The baseline essential and non-essential POIs visits were calculated based on the average trips from each CBG to essential POIs for the three weeks preceding Hurricane Harvey's landfall. The population was deemed recovered when CBG-to-POI visits reached 90% of baseline values for at least three consecutive days after landfall. The most rapidly recovering CBGs were recorded after one week, while the slowest recovering CBGs took more than 13 weeks to reach a level of movement to essential and non-essential POIs similar to pre-disaster levels Jiang et al. [54] details the methodology used to determine the baseline for the milestones.

### 2.2.1.2   CBG-to-CBG movements

The evacuation and move-out recovery milestones were calculated using Spectus data by looking at CBG-to-CBG movements based on the methodology proposed by Lee et al. [28].

*Evacuation recovery milestone* was calculated by considering the population who left their home CBG to travel to another CBG for at least 24 hours prior to Hurricane Harvey's landfall. The baseline was determined by computing the average of each weekday of the CBG-to-CBG movement between July 9, 2017, and August 5, 2017, inclusive. This average number was then divided by the total number of users in the census tract to obtain the baseline evacuation rate for each day of the week in each census tract.

*Move-out recovery milestone* is a measure assessing the home-switch or move-out rate, being a measure of how often people organically change residence and neighborhoods. This rate was computed by aggregating weekly data from July 9, 2017, through July 28, 2018. Each person in Harris County with a recorded home location was given a weekly home location tag. If no information about a person's home was generated for a week, it was assumed they did not move and used their last known home location. The move-out rate for each census tract area was computed by dividing the number of people who moved homes in a week by the total number of people in that area. To understand how this rate changed after Hurricane Harvey, it was compared to the average rate before the disaster, from July 9, 2017, through August 12, 2017. This comparison helped us see if there were more or fewer people moving homes after the landfall.

An area was considered recovered when the percent changes of the evacuation and move-out rates returned to a steady state. A steady state was defined as having no substantial difference in terms of the percent changes, meaning the difference between the percent change of the evacuation and home move-out rates between two consecutive days or weeks was within a threshold of 10%.

Using the evacuation rate percentage change, we considered an area to be recovered if the evacuation rate showed a variation of 10% from the baseline, lasted for at least three days after the landfall, and occurred before the Thanksgiving holiday. For more information, please refer to Lee et al. [28] to determine these milestones.

### 2.2.2   Physical vulnerability: Property damage extent indicator

To account for the physical vulnerability of spatial areas caused by flood events, flood claim data from the National Flood Insurance Program (NFIP) and the Individual Assistance (IA) program are utilized as a principal measure of residential property flood damage. This type of data provides critical insights into the extent of flood-related damage. The property damage extent (PDE) indicator, developed based on concepts proposed in scholarly literature [51, 56], measures the relative damage incurred by individual properties. The NFIP specifically insures against flood-related damage to buildings and contents, a requirement for residences in high-risk flood zones with federally backed mortgages. the Federal Emergency Management Agency's IA program offers financial assistance to disaster-impacted individuals for essential, uninsured, or under-insured expenses, though it does not aim to fully compensate for losses. The analysis focuses exclusively on structural damage using NFIP building claim payments and FEMA's real property damage assessments from the IA program. To ensure accuracy, only records with non-zero damage values from both sources are included, as zero values would indicate damage unrelated to flooding. For a comprehensive evaluation, NFIP and IA data are integrated using advanced geographical analysis techniques, mapping claim locations to specific building polygons with high geographic accuracy, primarily utilizing NFIP data due to its greater relevance over IA claims. The dataset includes 41,606 NFIP and 66,366 IA records from Harris County, Texas. The analysis resulted in the property damage equation, defined as:

$$PDE = \frac{building\ damage\ amount_i}{builidng\ property\ value_i}, \tag{1}$$

This metric aids in including damage relative to the property's value, providing a normalized scale for comparing damage across different property values. To harmonize property damage values from both datasets, a capping feature is applied at $500,000 for NFIP and $50,000 for IA, followed by min-max normalization. Consequently, the PDE indicator serves as an effective physical vulnerability proxy, offering a good depiction of the property damage caused by Hurricane Harvey.

### 2.2.3   Social vulnerability: Median Household Income

The median household income feature for 2017, extracted from the US Census Bureau American Community Survey [57]. used as a proxy of social vulnerability of spatial areas.

## 3   Methods

The milestones significant to represent population activity community recovery include evacuation recovery, recovery of essential and non-essential activities, and move-out recovery. Physical vulnerability is quantified using the PDE indicator, and social vulnerability is assessed through median household income. The representation of this framework is presented in Figure 1.

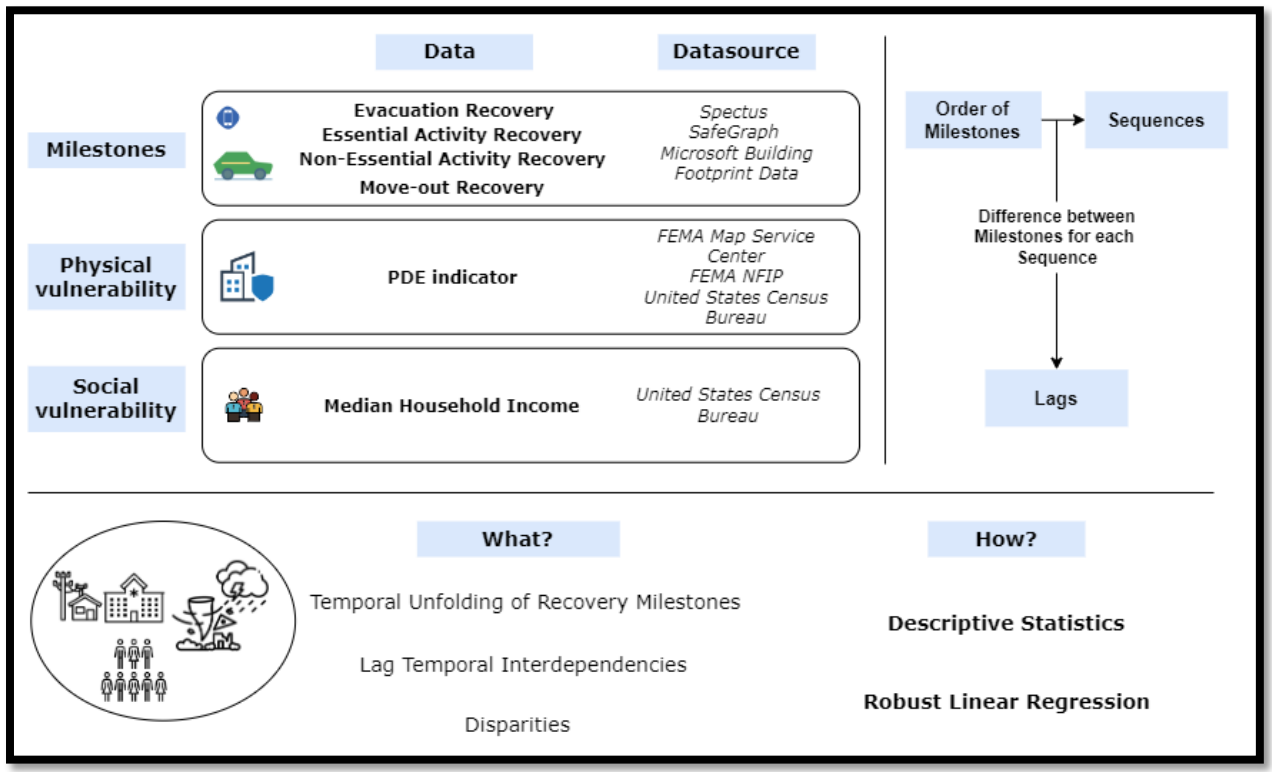

*Figure 1 Conceptual figure: Human mobility, property physical vulnerability damage indicator, and socio-demographic vulnerability information are considered to determine the variation of*

*the recovery milestone sequences via visitation to points of interests, property damage indicator, and median household income.*

## 3.1   Sequences of milestones unfolding

After identifying the critical milestones related to population activities, we analyzed the sequence in which these milestones unfolded across different areas using descriptive statistics. We examined distinct sequences of critical recovery milestones, each representing the unique recovery trajectories in spatial areas (census tracts) in Harris County in the aftermath of Harvey. In this analysis, we specified each critical milestone at the census-tract level throughout the case study area. Based on this analysis, we identified sequences in which recovery milestones unfold and examined comparisons within and across milestones. The results are presented in the next section.

## 3.2   Temporal independencies among milestones

After identifying the sequences related to the unfolding of critical recovery milestones, we examined the temporal interdependencies among milestones within each sequence based on the lags among them. We measured temporal lags by recording the time elapsed from the occurrence of the first milestone to that of each subsequent one (Figure 2).  This analysis reveals insights into the sequence and duration of recovery phases and how each phase may influence the onset and progression of subsequent ones. Certain milestones were found to occur typically in close succession, suggesting temporal co-dependence, or if there is a significant delay between them, potential delays. The focus of this analysis was to evaluate whether shorter lags among the initial milestones are associated with faster realization of the subsequent milestones.

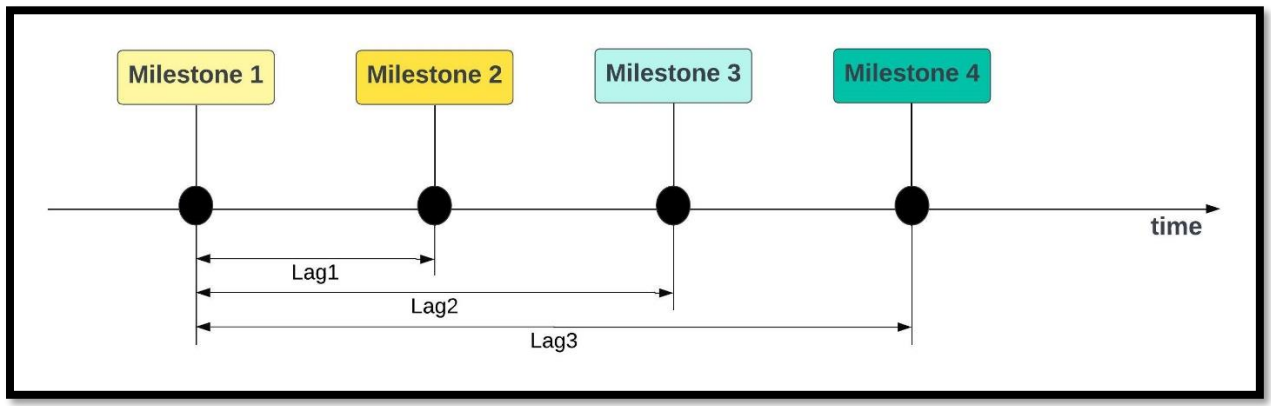

*Figure 2. The concept of Lags in Recovery Milestone*

To delve deeper into these temporal lags, we further employed a robust linear regression model. This advanced statistical method allowed us to quantify the interdependencies among the milestones, providing insights into the non-homogenous nature of the temporal unfolding of recovery. We opted for a robust linear regression model over a conventional one due to its effectiveness in handling data with non-normal distributions and outliers [58, 59], which are common characteristics of the recovery data from extreme weather-related events. Such data often challenge standard assumptions, presenting irregular patterns that require more flexible analytical approaches. By applying this model, we ensured our analysis remained accurate and reliable despite the inherent complexities of the data. The robust linear regression model we applied is described by the following equations:

$$Lag\ 2 = \beta_0 + \beta_1\ Lag\ 1 + \varepsilon \tag{2}$$

$$Lag\ 3 = \beta_0 + \beta_1\ Lag\ 2 + \varepsilon \tag{3}$$

In these equations, $\beta_0$ is the intercept, $\beta_1$ is the coefficient of the robust regression, and $\varepsilon$ represents the error term.

### 3.3   Within-sequence variations

We examined features of physical and social vulnerability to understand variations in the lag times among the critical milestones in each sequence. First, we selected the property damage extent indicator as a proxy for the extent of physical vulnerability of spatial areas resulting from a flooding event. Second, median household income was used as the indicator of social vulnerability. Although other indicators are relevant to the evaluation of social vulnerability, in this study, we only focused on household income since the prior literature showed that income levels play an important role in household recovery. We then explored the correlations among the lags between sequences, the PDE indicators, and median household income. To delve deeper into how physical and social vulnerability interact, we analyzed the distribution of median household income across each quantile of the PDE indicator. In addition, we examined how lags varied between the upper and lower segments of median household income relative to their median values. Further analysis was conducted to assess the differences in median household income associated with each lag within each sequence, focusing on the percentage deviations from the median value. The results of these analyses are presented in the following section.

## 4   Results and Discussion

### 4.1   Sequences of milestones unfolding

We identified six primary sequences of critical recovery milestones unfolding. This result suggests heterogeneity in the temporal unfolding of recovery milestones related to population activities. However, certain sequences dominate. The initial two sequences, encompassing 42.4% of the recorded areas, start with evacuation return recovery, followed by a variable order of essential and non-essential recovery milestones, and end with move-out recovery. Sequences 3 and 4 consistently begin with evacuation return recovery, succeeded by a flexible sequence of essential and move-out recovery, with non-essential recovery always concluding the sequence. The fifth and sixth sequences also initiate with evacuation recovery; however, they feature an interchangeable order of non-essential and move-out recovery, ultimately followed by the essential recovery phase. A minor segment, representing 3.82% of the census tracts, exhibited alternative recovery patterns. Due to their low prevalence, these atypical patterns were further excluded from the main analysis to maintain a focus on the predominant recovery trajectories.

*Table 1. Temporal unfolding of recovery milestones distribution*

| Sequence name | Sequence description | Sequence number of occurrences | Frequency |
|---|---|---|---|
| Seq1 | Evacuation recovery ≤ Essential activity recovery ≤ Non-essential activity recovery ≤ Move-out recovery | 183 | 23.28% |
| Seq2 | Evacuation recovery ≤ Non-essential activity recovery ≤ Essential activity recovery ≤ Move-out recovery | 150 | 19.08% |
| Seq3 | Evacuation recovery ≤ Essential activity recovery ≤ Move-out recovery ≤ Non-essential activity recovery | 141 | 17.94% |
| Seq4 | Evacuation recovery ≤ Move-out recovery ≤ Essential activity recovery ≤ Non-essential activity recovery | 133 | 16.92% |
| Seq5 | Evacuation recovery ≤ Non-essential activity recovery ≤ Move-out recovery ≤ Essential activity recovery | 77 | 9.80% |
| Seq6 | Evacuation recovery ≤ Move-out recovery ≤ Non-essential activity recovery ≤ Essential activity recovery | 72 | 9.16% |
| Other | Other sequences | 30 | 3.82% |

As shown in Table 2, sequence 1 recovers at average at a pace of 7.95 weeks (with move-out recovery, on average, last to be achieved), followed by sequence 2 at 8.49 weeks (with move-out recovery, on average, last to be achieved), sequence 5 at 8.63 weeks (with essential recovery, on average, last to be achieved), sequence 3 at 8.91 weeks (with non-essential recovery, on average, last to be achieved), sequence 6 at 9.29 weeks (with essential recovery, on average, last to be achieved), and sequence 4 at 9.70 weeks (with non-essential recovery, on average, last to be achieved). Table 2 presents the average duration of the final milestone for each sequence and their respective ranking based on this duration.

*Table 2. Descriptive statistics for sequences by recovery milestones*

| Sequence / Recovery Milestone | Evacuation | | Essential | | Non-essential | | Move-out | | Mean Maximum Duration | Rank |
|---|---|---|---|---|---|---|---|---|---|---|
| | Mean | St. Dev. | Mean | St. Dev. | Mean | St. Dev. | Mean | St. Dev. | | |
| Seq1 | 0.85 | 0.35 | 2.79 | 1.70 | 4.76 | 2.49 | 7.95 | 2.75 | 7.95 | 1 |
| Seq2 | 0.83 | 0.37 | 5.27 | 2.43 | 3.43 | 1.78 | 8.49 | 2.93 | 8.49 | 2 |
| Seq3 | 0.85 | 0.39 | 3.73 | 2.13 | 8.91 | 2.62 | 6.12 | 2.12 | 8.91 | 4 |
| Seq4 | 0.75 | 0.35 | 7.18 | 3.37 | 9.70 | 3.10 | 4.08 | 2.64 | 9.70 | 6 |
| Seq5 | 0.82 | 0.42 | 8.63 | 2.62 | 3.97 | 2.24 | 6.52 | 2.23 | 8.63 | 3 |
| Seq6 | 0.72 | 0.28 | 9.29 | 2.72 | 6.77 | 2.15 | 4.17 | 2.19 | 9.29 | 5 |

Moreover, the results indicate significant variability in recovery times across different sequences and recovery types. For evacuation recovery, sequences exhibit relatively small standard deviations, suggesting consistent duration for these critical milestones, with sequence 6 showing the least variability (0.28). Sequence 4 stands out with the highest standard deviation for both essential (3.37) and non-essential recovery (3.10), indicating greater heterogeneity. Move-out recovery also shows notable variability, with sequence 2 having the lowest standard deviation (1.78) and sequence 2 the highest (2.93). Sequences with lower standard deviations, such as sequence 6 for evacuation and move-out recovery, imply more predictable and consistent recovery timelines. Conversely, sequences like sequence 4 for essential and non-essential recovery suggest less predictability and more dispersed recovery times. These results show the heterogenous nature of duration for different recovery milestones suggesting the effects of factors such as social and physical vulnerability which will be discussed in the rest of the results.

Understanding the variability in recovery times is fundamental for effective recovery planning and resource allocation. People and physical assets within sequences with higher standard deviations may require more tailored strategies to address their unique challenges effectively. The chronological order of recovery milestones, as illustrated in Table 2 and Figure S1, provides insights into the dynamics of recovery, highlighting both dominant and less common patterns. This nuanced analysis provides actionable insights for emergency managers to enhance community resilience effectively. For instance, by identifying vulnerable neighborhoods which follow sequence 4 for recovery with, on average, the most delayed recovery times, managers can prioritize resource allocation and support services in that area. Monitoring recovery patterns can also highlight areas where interventions, such as infrastructure upgrades or community outreach programs, are most needed. For instance, if an area's recovery pattern follows sequence 6, emergency managers can proactively prioritize essential resource allocation within that area, recognizing that essential services are expected to recover last in this sequence. These targeted approaches not only optimize recovery outcomes but also ensure that support efforts are tailored to meet the unique challenges of each community, fostering long-term resilience.

## 4.2   Temporal independencies among milestones

Figure 3 illustrates the duration of lags for each recovery sequence, revealing a notable trend. Specifically, the mean lag values indicate a progressive increase in the duration of the lags from sequence 1 to sequence 6. This pattern demonstrates that the initial lag in sequence 1 occurs more swiftly compared to sequence 6, where the onset is significantly delayed. Also, the data suggest a marked temporal interdependence among the sequences. Shorter initial lags (lag 1) are coupled with shorter durations in subsequent lags (lag 2 and lag 3). This observation indicates a ripple effect: a faster start in the recovery process can lead to an accelerated overall recovery. Enhancements in the initial stages of recovery could positively impact the entire recovery timeline, as prompt initial actions post-disaster can have a cascading positive effect.

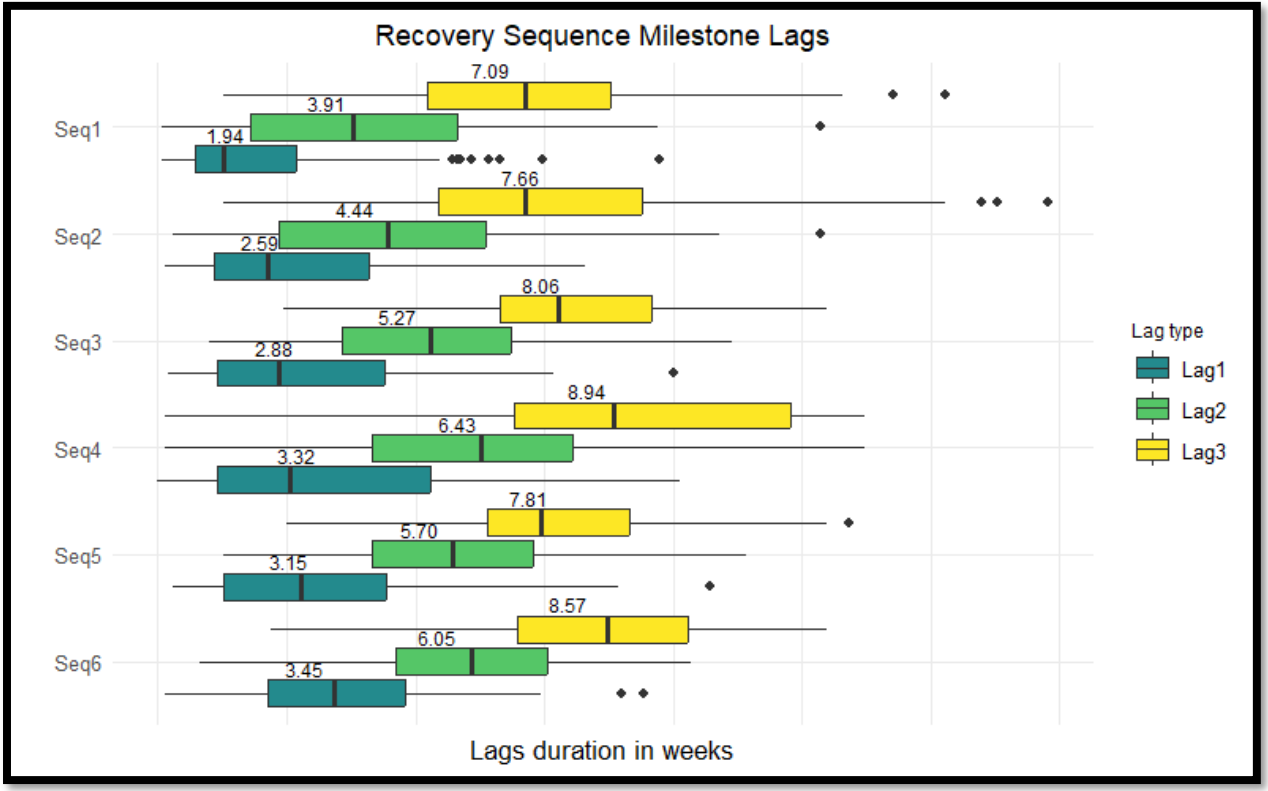

*Figure 3. Results of lags between milestones for each sequence*

To validate the statistical results presented in Table 2 and the visual observations from Figure 3, a robust linear regression model was employed. The detailed findings from this analysis are summarized in Table 3. The analysis confirms significant relationships between the lags in each sequence, all with a p-value less than 0.001. Specifically, lag 1 significantly correlates to lag 2 across all sequences, with the explanatory power ranging from 0.760 in sequence 2 to 0.379 in sequence 6. Similarly, lag 2 has implications over lag 3, with the strength of this relationship varying from 0.728 in sequence 5 to 0.402 in sequence 2. While the visual analysis indicated a general increase in lag durations from sequence 1 to sequence 6, the robust regression model reveals more nuanced correlations: there is a strong correlation between lag 1 and lag 2 in sequences 1, 2, and 4, and a moderate one in sequences 3, 5, and 6. Conversely, the correlation between lag 2 and lag 3 is strong in sequences 3, 5, and 6, and moderate in sequences 1, 2, and 4. These findings underscore the importance of examining and monitoring the temporal interdependencies among various critical recovery milestones. By identifying and understanding these relationships, emergency managers and public officials can proactively identify areas with slower initial recovery milestones which would cascade to delays in the subsequent milestones.

*Table 3. Robust linear regression model per sequence*

| | Dependent variable: | | | | | | | | | | |
|---|---|---|---|---|---|---|---|---|---|---|---|
| | Lag2 | | | | | | Lag3 | | | | | |
| | Lag2 | Lag2 | Lag2 | Lag2 | Lag2 | Lag2 | Lag3 | Lag3 | Lag3 | Lag3 | Lag3 | Lag3 |
| | (1) | (2) | (3) | (4) | (5) | (6) | (7) | (8) | (9) | (10) | (11) | (12) |
| Lag1 | 0.732*** | 0.760*** | 0.464*** | 0.601*** | 0.473*** | 0.379*** | | | | | | |
| | (0.064) | (0.055) | (0.051) | (0.061) | (0.071) | (0.080) | | | | | | |
| Lag2 | | | | | | | 0.496*** | 0.402*** | 0.600*** | 0.557*** | 0.728*** | 0.668*** |
| | | | | | | | (0.049) | (0.058) | (0.063) | (0.045) | (0.060) | (0.082) |
| Constant | 0.116*** | 0.106*** | 0.214*** | 0.210*** | 0.218*** | 0.257*** | 0.212*** | 0.239*** | 0.181*** | 0.199*** | 0.109*** | 0.154*** |
| | (0.013) | (0.013) | (0.014) | (0.019) | (0.020) | (0.025) | (0.014) | (0.017) | (0.021) | (0.018) | (0.021) | (0.031) |
| Seq | Seq1 | Seq2 | Seq3 | Seq4 | Seq5 | Seq6 | Seq1 | Seq2 | Seq3 | Seq4 | Seq5 | Seq6 |

**Robust Linear Regression Models for Each Seq Category**

*Note:* $^{*}p<0.1$; $^{**}p<0.05$; $^{***}p<0.01$

The implications of these findings are threefold. First, investments in rapid response capabilities could yield significant benefits throughout the recovery process, particularly affecting later lags. Proactive recovery efforts at the sequential level can lead to more synchronized and effective progression through the recovery milestones, enhancing resilience against potential setbacks. Second, understanding that the duration of early recovery activities impacts subsequent stages can guide targeted effective resource allocation. Emergency managers and agencies could prioritize efforts and resources to trigger the initial recovery milestones as quickly as possible, thereby enhancing the overall speed of recovery. Finally, these findings contribute to the advancement of knowledge in the disaster risk management field by providing empirical evidence of the criticality of considering temporal interdependence in disaster recovery studies.

## 4.3 Within-sequence variations

The results in Figure 4 depict the variation in recovery sequences across different quantiles of the PDE indicator and the Median Household Income. The solid-filled sequences depict the quantiles, including Q1, Q2, Q3, and Q4, of the median household income and PDE indicator quantiles Q1 and Q4. These quantiles segment the data into four parts, each representing different income levels. Meanwhile, the dashed line shows the distribution of frequencies of these sequences divided into four equal parts. This division normalizes the frequency across the four quantiles, facilitating easier comparison of the data distribution across different segments. The results suggest that areas within the lower quantiles of both PDE indicator and household income predominantly follow recovery trajectories aligned with sequence 4. Conversely, areas in the upper quantiles for both metrics adhere to sequences 1 and 2. This variation indicates that social and physical vulnerability factors explain variations in the sequence of recovery milestones. As shown in the previous results, however, the variation in the lags among milestones within each sequence is primarily explained by social vulnerability, with lower-income areas experiencing longer lags.

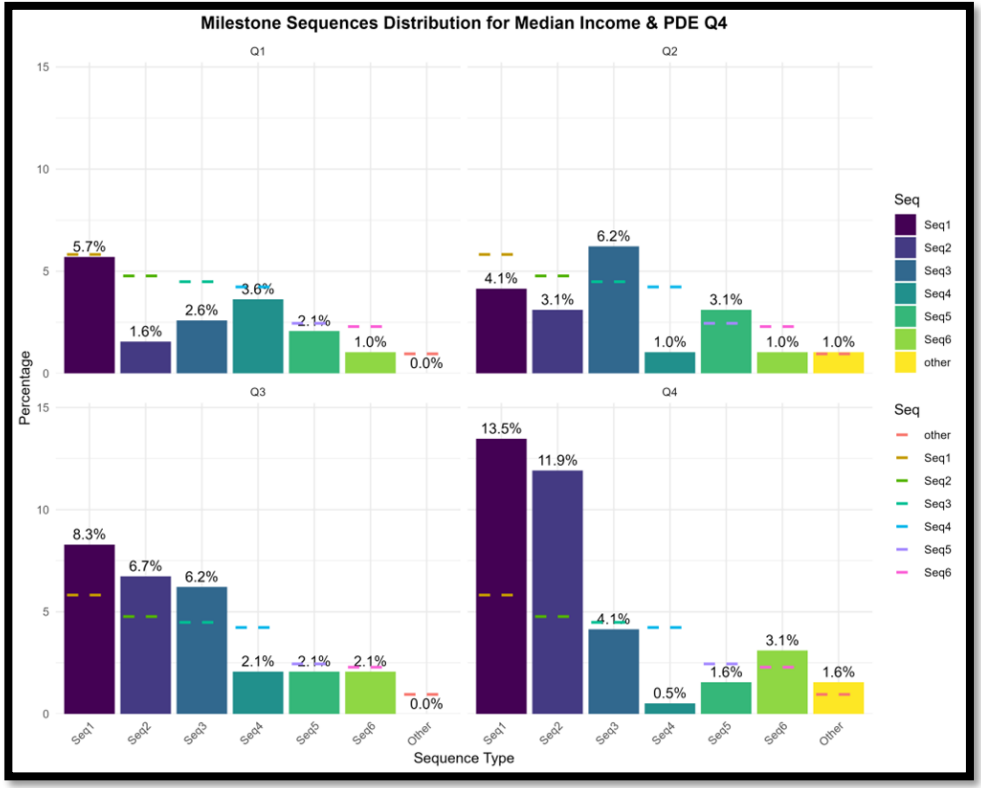

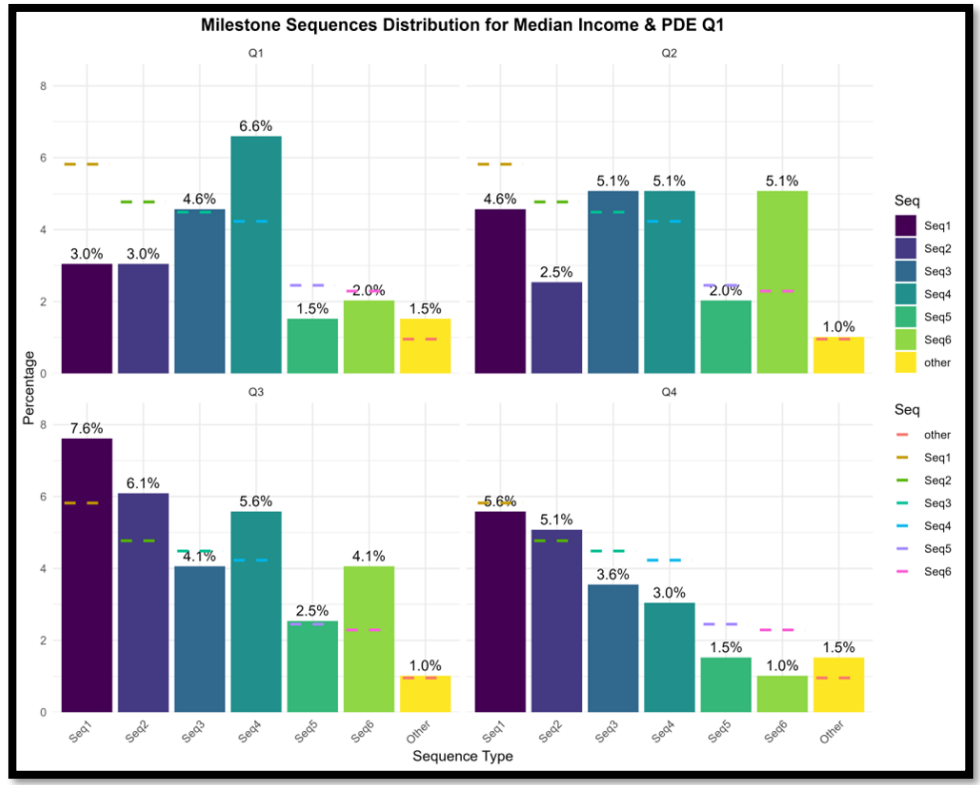

*Figure 4. (a) Household median income for PDE Q1 sequential unfolding (b) Household median income for PDE Q4 sequential unfolding*

Further, we examined the variation in the lags among the milestones within each sequence based on physical and social vulnerability indicators of the areas as shown in Figure 5. Our analysis revealed a significant correlation between the milestone lags and median household income, whereas the correlation between the lags and the PDE indicator was not deemed significant. This finding suggests that variations in sequences and lags are more closely associated with social, rather than physical vulnerability.

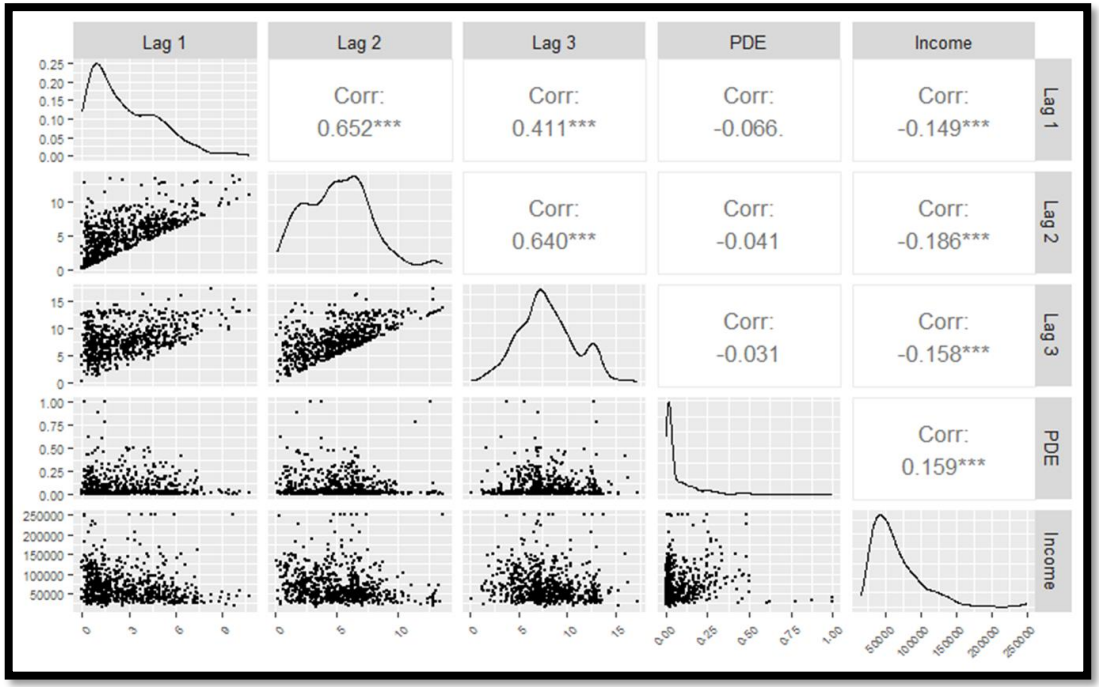

*Figure 5. Relationship between lag 1, lag 2, lag 3 , PDE indicator and household median income*

We continued our analysis by examining the disparities in recovery sequences relative to median household income, focusing specifically on the upper- and lower-income quantiles. The results, illustrated in Figure 6, reveal that populations in lower-income areas generally experience significantly longer recovery periods compared with their higher-income counterparts. These findings underscore the profound impact of economic disparity on community recovery rates. This disparity highlights the role of socioeconomic factors in community recovery trajectories to disasters. Populations with higher incomes are likely to have better access to resources that can expedite their recovery—such as insurance, savings, or credit—which are less accessible to lower-income groups. Understanding these dynamics is essential for crafting more equitable disaster response strategies that address these imbalances.

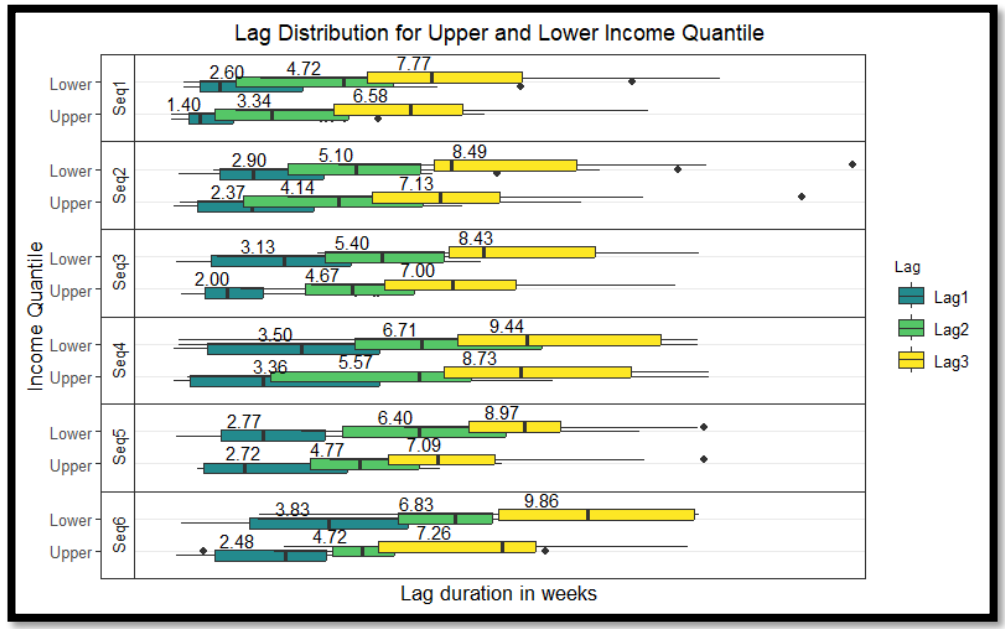

*Figure 6. Upper and lower quantile median household income variation by recovery trajectory sequence*

Building on the previous findings, we further analyzed the extent to which income levels at the upper and lower quantiles deviate from the mean for each lag type across the sequences (Figure 7). In particular, Figure S2 generated using Table 4 illustrates the percentage variations for upper-, median-, and lower-income quantiles across different recovery trajectory sequences and lags. The results reveal significant variations in recovery speeds influenced by income levels, which manifest distinctly across all sequences in a non-homogeneous manner. For instance, sequence 1, in lag 1, lower-income groups, recovers 33.74% slower compared to the mean, while higher-income groups recover 28.09% faster. This difference decreases in lag 2 to 20.41% slower for lower-income groups and 14.87% faster for higher-income groups, and further narrows in lag 3 to -7.38% and 9.42%, respectively. In contrast, for sequence 4 (last one to recover) recovery dynamics vary significantly: in the lower quantile for income, recovery is on average slower by 5.64%, 5.16%, and 5.90% for lag 1, lag 2, and lag 3 compared to the mean value, while the upper quantile shows a slower recovery for lag 1 of 1.57% compared to the mean and a faster recovery of 12.65% and 2.04% for lag 2 and lag 3, respectively.

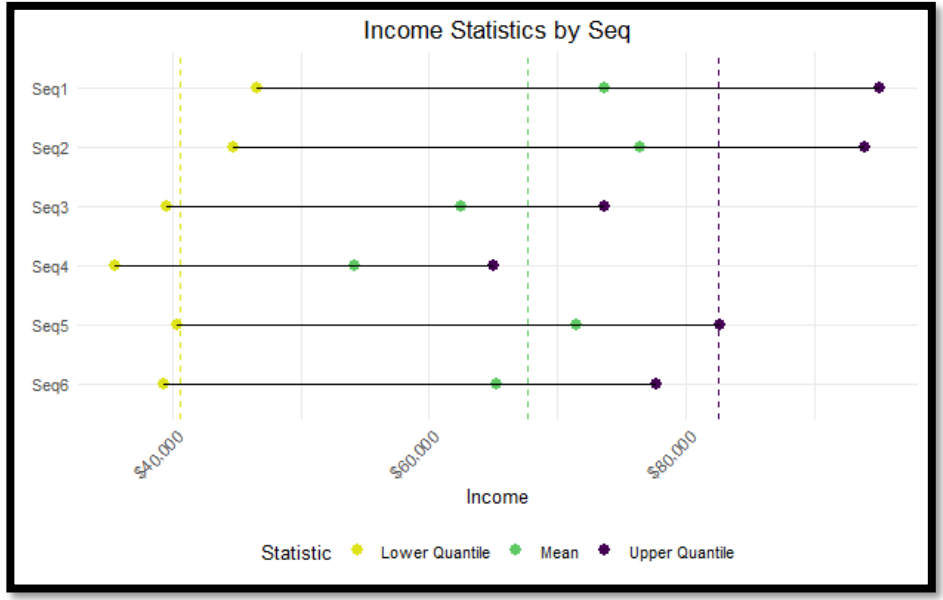

*Figure 7. Upper, median, and lower household income quantile per recovery trajectory sequence*

The results presented in Table 4 reinforce the idea that targeted intervention early in the recovery process for lower-income groups has the potential to reduce initial disparities and foster a more equitable recovery across socioeconomic groups. Such an approach not only addresses immediate recovery needs but also contributes to long-term community resilience by ensuring that all segments of the population recover more uniformly. By implementing such strategies, policymakers can work towards building more inclusive and resilient communities in the aftermath of disasters.

*Table 4. Percentage Changes for Upper and Lower Quantile Household Income*

| Name | Quantile | % Change Lag 1 | % Change Lag 2 | % Change Lag 3 |
|------|----------|----------------|----------------|----------------|
| Seq1 | Upper | -28.09 | -14.87 | -7.38 |
|      | Lower | 33.74 | 20.41 | 9.42 |
| Seq2 | Upper | -8.49 | -6.79 | -6.82 |
|      | Lower | 11.71 | 14.86 | 10.83 |
| Seq3 | Upper | -30.63 | -11.42 | -13.16 |
|      | Lower | 8.67 | 2.32 | 4.53 |
| Seq4 | Upper | 1.57 | -12.65 | -2.04 |
|      | Lower | 5.64 | 5.16 | 5.90 |
| Seq5 | Upper | -13.56 | -16.20 | -9.20 |
|      | Lower | -11.90 | 12.30 | 14.86 |
| Seq6 | Upper | -28.02 | -22.00 | -15.25 |
|      | Lower | 11.20 | 12.94 | 15.06 |

These insights underscore the critical role of socio-economic factors in disaster recovery management, emphasizing the need for targeted interventions to address proven inequalities in recovery. Understanding these dynamics empowers policymakers and disaster management professionals to allocate resources more effectively, promoting long-term community resilience by prioritizing vulnerable areas. Integrating targeted support mechanisms into recovery plans ensures that vulnerable populations efficiently restore their livelihoods and communities in the short term, accelerating both short- and long-term recovery efforts while bolstering societal resilience against future disasters. In practice, emergency managers should prioritize targeted interventions, for example, for areas which are among the slowest to recover (sequences 5 and 6) with the essential services milestone being the last to be achieved. Additionally, areas with the lowest household income (sequence 4) similarly require focused attention. For longer-term recovery, planners can promote new investments in essential and nonessential facilities to achieve a balanced distribution of facilities for the above-mentioned sequences. This would reduce the overall travel time to reach these facilities and thereby accelerate recovery in the aftermath of a disaster.

# 5   Concluding Remarks

Understanding the unfolding of disaster recovery milestones is essential for effective short-term recovery and long-term resilience. This study addresses the imperative for immediate, coordinated responses following disasters, crucial for initiating recovery efforts, mitigating further harm, and addressing the needs of the population. By examining the temporal progression of recovery milestones across different sequences, the study elucidates their associated dynamic interdependencies and time-varying patterns. Additionally, this study investigates disparities in recovery sequences linked to both physical and social vulnerability, using the property damage extent indicators and median household income. The overarching outcome of implementing such findings is to enhance disaster response efforts in the aftermath of weather-related disasters and strengthen long-term community resilience.

The study analyzed evacuation, essential activity, non-essential activity, and move-out recovery milestones, identifying six predominant sequences in which these milestones unfold. Empirical data from Harris County in 2017, during Hurricane Harvey, informed the computation of community recovery milestones. These were derived from human mobility data, property damage indicators (based on NFIP claims and IA assistance for flood-affected areas, reflecting physical vulnerability), and household median income (reflecting social vulnerability). Recovery patterns exhibited significant variation across different areas, influenced by local lifestyle choices, although this aspect falls beyond the study's scope.

The study identifies and characterizes the distinct recovery timelines across sequences. Sequence 1 emerges as the fastest to recover, averaging 7.95 weeks, followed closely by sequence 2 at 8.49 weeks, and sequence 5 at 8.63 weeks. The last sequence to recover was sequence 4 at an average of 9.70 weeks. The robust linear regression has identified interdependencies between lags meaning that the longer initial recovery lags between the first and second milestones lead to even higher delays between subsequent milestones, both overall and within each sequence. When looking at the variation in both in income distribution and property damage extent indicator, lower-income populations with lower property damage extent (lower quantile PDE) tend to follow sequence 4 (the slowest recovery sequence identified), where evacuation recovery is succeeded by move-out recovery, followed by the recovery of essential and non-essential services. Conversely, higher-income populations with lower property damage (upper quantile PDE) typically follow sequences 1 and 2 (the two fastest recovery sequences identified), where evacuation recovery is followed by the recovery of essential and non-essential services, with move-out recovery occurring last. This suggests that lower-income groups may prioritize moving out due to limited resources, while higher-income groups can afford to follow a more standard recovery sequence facilitated by their financial stability. Our analysis revealed a significant correlation between milestone lags and median household income, but not with the PDE indicator. This suggests that variations in sequences and lags are more closely associated with social vulnerability than physical vulnerability. When looking at the disparities, for example, sequence 1, the fastest to recover, shows significant income disparities: lower-income groups lag 33.74% behind the mean, while higher-income groups recover 28.09% faster. These gaps decrease in lag 2 to 20.41% slower for lower-income and 14.87% faster for higher-income groups, further narrowing in lag 3 to -7.38% and 9.42%, respectively. Conversely, sequence 4, the slowest to recover, reveals different patterns: upper-income groups recover 1.57% slower than the mean in lag 1 but faster by 12.65% and 2.04% in lags 2 and 3. Lower-income groups recover slower than mean in all lags with 5.64%, 5.16%, and 5.90% in lags 1, 2, and 3, respectively.

These insights underscore the necessity for targeted interventions in disaster response planning. Understanding variations in recovery sequences linked to socio-economic status and property damage enables policymakers to allocate resources effectively and develop tailored strategies for vulnerable populations. For example, regardless of physical flood damage vulnerability, lower-income groups experience significantly longer recovery times compared to higher-income counterparts, highlighting the critical role of socio-economic vulnerability in recovery dynamics. The empirical identification of these recovery patterns provides a framework for predicting future disaster recoveries, thereby enhancing preparedness and disaster management strategies. Furthermore, the interdependencies between the recovery lags provide evidence of a ripple effect in community recovery: a faster start in the recovery process can lead to an accelerated overall recovery. This study advances the field by empirically demonstrating the impact of both physical and social vulnerability factors on recovery processes and proposes strategies to address equitable recovery issues in both short-term recovery and long-term community resilience planning.

Finally, it is crucial to acknowledge certain limitations in the study's methodology. Primarily, data collection depends on smartphone usage, which only includes users who have consented to location tracking. This method may omit significant demographic groups such as children, teenagers, the elderly, and lower-income individuals, thereby potentially skewing the data and introducing biases [60, 61]. Nevertheless, to address these concerns partially, we incorporated Spectus data, known to represent a diverse cross-section of the population [32].

## Funding

The author(s) disclosed receipt of the following financial support for the research, authorship, and/or publication of this article: National Science Foundation number CMMI-1846069 (CAREER).